\documentstyle[11pt,paspconf]{article}

\begin{document}

\title{Baryon Crisis and Cluster Mass Estimates}

\author{Xiang-Ping Wu}
\affil{Institute of Astronomy, National Central University, Chung-Li
                 Taiwan 32054, ROC; and
      Beijing Astronomical Observatory, Chinese Academy
                 of Sciences, Beijing 100080, China}

\begin{abstract}
I report the current status on the determinations of the
baryon fractions and dynamical/lensing masses of galaxy clusters
as well as the $\sigma$ - $T$ relationship,
making use of all the published data in literature 
which include 304 measurements for baryon fractions, 
320/21 data points for masses of 
intracluster gas/cluster galaxies, (152+228)/72 data points for cluster
masses derived from X-ray emitting gas/optical galaxies based on
the hydrostatic equilibrium hypothesis, 46/33 measurements for
the projected cluster masses obtained from strong/weak lensing. 
A straightforward statistical analysis yields the following
features: (1)The cluster baryon fraction indeed shows an
increasing tendency to reach a value of 
$\sim30\%$ at large cluster radius. 
(2)The dynamical cluster masses are in fairly
good agreement with the masses determined by 
weak lensing technique; (3)The cluster
masses provided by strong lensing and other methods
are essentially consistent within a factor of $\sim2$.
(4)The cluster $\sigma-T$ relationship is in concordance with
the scenario that overall galaxy clusters are the well
relaxed dynamical systems. These results are strongly suggestive of 
a low-mass density universe ($\Omega_m<0.3$), in which the fraction of
baryonic matter ($\sim30\%$) is not very small ! 
\end{abstract}


\section{Introduction}

The standard Big Bang Nucleosynthesis (BBN) has set rather a tight
limit on the baryonic matter density of the universe
(Walker et al. 1991; Schramm \& Turner 1997):                      
$0.028<\Omega_b<0.08$ (I adopt $H_0=50$ km s$^{-1}$ Mpc$^{-1}$
throughout). 
The recent measurements of various baryonic matter components 
in the nearby universe indeed yield a good agreement with the 
BBN prediction (Fukugita et al. 1997),
indicative of a relatively small fraction of baryonic matter 
if the total mass density of the universe is
$\Omega_m>0.1$, i.e., a significant fraction of the mass in the universe
is invisible.  With the progress in X-ray technique over the past
decade,  a considerably large amount of hot X-ray emitting gas has been
detected in groups and clusters of galaxies, which gives rise to a baryon
fraction of up to $f_b\sim30\%$ (see Figure~\ref{fig-1}). 
If clusters are a fair
sample of matter composition of the universe and if a flat cosmological
model of $\Omega_0=1$ is acceptable, we are inevitably faced with the
difficulty of the so-called baryon crisis  (White et al. 1993).               
The most prevailing solution to the puzzle today is to work with   
an open universe with matter density $\Omega_m<0.3$,               
which is, nevertheless, consistent with the result                 
obtained by a number of recent                                     
studies on the abundances and evolution of galaxy clusters,        
in particular the mass-to-light ratio $M/L$ measurements           
(e.g. Bahcall \& Cen 1993;                                         
Bahcall et al. 1995; Carlberg et al. 1997; Bahcall et al. 1997).

The {\it key} issue of either confirming or resolving the baryon crisis
is closely connected to the question of how
accurately one can determine the masses of galaxy clusters.
Recall that the traditional cluster estimators strongly rely upon the
assumption of hydrostatic equilibrium: both optical galaxies and
intracluster diffuse
gas trace the underlying gravitational potential of the whole cluster.
An independent method of estimating cluster mass, which has been
available only for decade, is to employ the 
gravitational lensing technique. It gives rise to cluster
masses regardless of the cluster matter components 
and their dynamical states.  A series of work has thus been made on
a statistical comparison of the cluster masses derived
from the traditional dynamical method and the gravitational
lensing based on the published data in literature
(Wu 1994; Wu \& Fang 1996, 1997; Wu et al. 1998a).
In this talk, I summarize the most recent finding of such series. 
The purpose is to demonstrate how large the uncertainties could be      
among the present various cluster mass estimates.                      
Such a statistical comparison may eventually help towards 
confirming or resolving the baryon crisis
and the $\Omega_0$ disprepancy, if 
a flat universe is preferred by the standard inflationary
model, and a better understanding of the dynamical evolution
of galaxy clusters.

\section{Sample}

In order to avoid prejudices that may have arisen from our
selection criteria, we adopt only the published data sets in literature and
make no attempt to extrapolate the original work.
For example, we will not make additional computation of the
X-ray cluster masses that were not provided by the authors,
even if we have known the gas density and temperature distributions.
Based on this strategy,
an extensive search for literature yields 21 measurements for galaxy
masses in clusters, 320 data for mass of intracluster gas,
79 data for virial cluster masses derived from optical galaxies,
152+226 data for X-ray cluster masses derived from X-ray emitting gas,
8 data for X-ray group masses, and 46/33 data for the projected cluster
masses derived from strong/weak lensing. Compared to
the cluster sample of White, Jones and Forman (1997 
and hereafter WJF), the present sample extends
the data of baryon fractions from 176 to 304, 
the X-ray cluster masses from 226 to 378, the gas masses from 226
to 320. Note, however, that the data points are apparently larger than
the actual cluster population because in many cases 
several measurements have been carried out toward a single cluster. 
We have simply added the observed/derived data that are not among
the WJF sample.  Our data sets and references will be
reported elsewhere (also available upon request).

\section{Baryon fraction}

Figure~\ref{fig-1} displays the 304 measurements of cluster baryon (gas) 
fractions $f_b$ for 238 clusters obtained at different radii. 
With the data being properly binned according to radius, 
an increasing tendency of $f_b$ along cluster radius
is clearly presented. Such a variation was actually noticed 
two decades ago and has been confirmed by a number of
recent observations (e.g. White \& Fabian 1995; 
Ettori et al. 1997; David 1997; WJF).  
Two conclusions can be drawn immediately: (1)Because
the cluster baryon fraction is representative of
the universal value, the measurements of the
baryon fractions over an ensemble of
clusters shown in Figure~\ref{fig-1} 
thus imply a cosmological value of 
$f_b=\Omega_b/\Omega_m\approx30\%$. 
This leads to  $\Omega_m<0.3$ 
combined with the BBN prediction of $\Omega_b<0.08$. 
(2)The increase of baryon fraction with radius 
indicates that a greater fraction of dark matter is 
distributed at small scales than at large scales,
in contradiction with the conventional point of view.

\begin{figure}                            
\vspace{3.75in}                           
\caption{(a)The baryon fractions in clusters  of galaxies
obtained from  X-ray observations under the assumption of
isothermal and hydrostatic equilibrium. A total of 304 data points
for 238 clusters in literature are shown. (b)The same as (a) but
the data sets are binned such that each bin contains 19 measurements.
The observed baryon fractions in galaxies and groups of galaxies are also
illustrated.    The dashed lines show the BBN prediction
for a flat universe of $\Omega_0=1$,  for which
we have already accounted for the uncertainty in the recent determination
of the primordial Deuterium abundance. }
\label{fig-1}                             
\end{figure}

\section{Statistical comparison of mass estimates}                                   

The original data of the observationally determined cluster masses versus         
cluster radii are shown in Figure~\ref{fig-2}, in which we have also
displayed the projected mean cluster mass required to
preduce the quasar-cluster associations
detected at the smallest cluster radius $r=1.5$ Mpc among the
four measurements (Wu \& Fang 1996a), and
the recent result based on the red galaxy counts behind A1689
(Taylor et al. 1998).

\begin{figure}
\vspace{2.5in}
\caption{Different mass estimates versus radii for galaxy clusters.
All the data points are taken from literature.
Note that the cluster masses derived from gravitational lensing
are the projected ones.}
\label{fig-2}
\end{figure}

The X-ray/optical selected clusters exhibit a large dispersion 
in their dynamical masses due to the difference of richness.
On the other hand, the lensing clusters are usually the most  
massive ones at intermediate redshifts, which enables them 
to act as lenses for distant galaxies.
This requires that our comparison of different cluster mass
estimates can only be made among the hot and massive clusters.
Examination of the lensing cluster sample (see also  Wu \& Fang
1997) shows that the mean X-ray temperature and 
galaxy velocity dispersion  of lensing clusters are
approximately $T=7.5$ keV and  $\sigma=1200$ km s$^{-1}$,
respectively.  We slightly relax these limits and 
use the criteria of $T\geq7$ keV or $\sigma\geq1100$ km s$^{-1}$ 
to select the X-ray/optical clusters for comparison.
This leaves us 49/56 measurements from the WJF/other samples
and 18 data for the virial cluster masses $M_{vir}$. 
Furthermore, in order to facilitate a statistical comparison 
among different cluster mass estimates, 
the selected data sets are properly binned and shown in 
Figure~\ref{fig-3}.  We fit the X-ray cluster mass $M_{xray}$ 
using  an isothermal $\beta$ model               
with core radius $r_c$ for the distributions of hot X-ray emitting 
gas based on hydrostatic equilibrium hypothesis
(Figure~\ref{fig-3}).                    
Our best-fit of the X-ray data (WJF + others) for $\beta=2/3$ 
to $M_{xray}=2(kT/G\mu m_p)r^3/(r^2+r_c^2)$  
gives $r_c=0.25$ Mpc and $T=9.5$ keV. 
Apparently, no significant      
difference between the mean values of $M_{xray}$ and $M_{vir}$ is detected.
Since the gravitational lensing always provides a projected gravitating
cluster mass $m_{lens}$, we need to transform                      
the best-fitted mass profiles $M_{xray}$ into the corresponding 2-D masses
$m_{xray}$ for the purpose of comparison. Alternatively,                                         
it should be noted that the current weak lensing technique only sets a lower
bound on $m_{lens}(r)$ within cluster radius $r$ (e.g. Fahlman et al. 1994):
$m_{lens}(r)>\pi r^2\Sigma_{crit}\int_{r}^{r_{max}}\langle\gamma_T\rangle
(1-r^2/r_{max}^2)^{-1}d\ln r$, where                               
$\Sigma_{crit}=(c^2/4\pi G)(D_s/D_dD_{ds})$ is the critical mass density,
with $D_d$, $D_s$ and $D_{ds}$ being the angular diameter distances to the
cluster, to the background galaxies, and from the cluster to the galaxies,
respectively, $\langle\gamma_T\rangle$                             
denotes the shear effect on the image configuration of background galaxies
introduced by the intervening cluster gravitational potential, and 
$r_{max}$ is the maximum radius of a control annulus.

\begin{figure}
\vspace{3.5in}
\caption{(a)Statistical comparison of cluster masses inside radii $r$
derived from  dynamical analysis and gravitational lensing.
We only select the massive X-ray/optical clusters in Figure~\ref{fig-2} 
that satisfy $T\geq7$ keV or $\sigma\geq1000$ km s$^{-1}$. The
dynamical masses for these selected clusters
are properly binned for illustration and fitting.
The dashed/solid lines are the best-fit of the 3-D/2-D
X-ray cluster masses to the theoretically expected results
under the hydrostatic equilibrium hypothesis for an isothermal
$\beta$ gas distribution with $r_c=0.25$ and $T=9.5$.
Multiplying the best-fitted 2-D X-ray cluster mass (solid line)
by a factor of 2 gives the result shown by dotted line.
The binned gas mass distribution of all the measurements (320) 
is also shown.}
\label{fig-3}
\end{figure}

It turns out that the dynamical masses of clusters  given by
the traditional methods,
namely X-ray analysis and virial theorem,  are
in good agreement with the gravitating cluster
masses revealed by weak lensing.
While we cannot exclude the possibility that 
dynamical analysis may systematically underestimate 
the true cluster masses as compared with strong
lensing, the mass discrepancy, if any, is well within a 
factor of $\sim2$. When the measurements errors are included
and the uncertainties in our selection criteria 
of $T\geq7$ keV or $\sigma\geq1000$ km s$^{-1}$ are
taken into account, we conclude that there is no
significant mass discrepancy among the various
cluster mass estimators. This is compatible with
the previous similar work made for individual lensing
clusters (e.g. Allen et al. 1996; Squires et al. 1996; 
Markevitch 1997; Small et al. 1997; Allen 1997; 
Wu \& Fang 1996a,b,1997).

\section{The $\sigma$-$T$ relationship}   

The consistency between the different mass estimates indicates that
as a whole, galaxy clusters can be regarded as the well relaxed 
dynamical systems, though substructures and complex temperature patterns
have been observed at small scales. A simple and robust way to test 
this scenario is to employ the velocity dispersion - temperature
relationship: If both the galaxies and gas are the tracers of
the depth and shape of a common gravitational potential, we
would expect $\sigma\sim T^{0.5}$ (Cavaliere \& Fusco-Femiano 1976).
In Figure~\ref{fig-4}
we show the updated $\sigma$-$T$ relationship based on
141 clusters selected from our largest cluster sample, for which 
both $\sigma$ and $T$ are observationally determined, where
we exclude those clusters whose $\sigma$ and/or $T$ are obtained
by indirect methods such as the $L_x$ - $\sigma$ and
$L_x$ - $T$ correlations. Our best-fitted result reads
$\sigma=10^{2.57\pm0.03}T^{0.53\pm0.04}$ (unweighted) and
$\sigma=10^{2.53\pm0.02}T^{0.60\pm0.04}$ (weighted), which
are fully consistent with previous similar work but with a reduced
s.d. of the residuals (see Wu et al. 1998b and references
therein). Therefore, at  $3\sigma$ confidence level
the $\sigma$-$T$ relationship also supports for
cluster of galaxies being a virialized system.

\begin{figure}
\vspace{2.25in}
\caption{The $\sigma$-$T$ relationship for the 141 clusters
in our sample. The low-redshift ($z,0.1$) and high-redshift
($z\geq0.1$) clusters are represented by the open triangles
(106) and the filled squares (35), respectively. The solid
line is the (unweighted) best-fit to the data.}
\label{fig-4}
\end{figure}

\section{Discussion and conclusions}                               

The overall observed baryon fraction of galaxy clusters
is $\sim30\%$. This value has been justified by our  
statistical comparison of various mass estimates of clusters published
so far in literature, which shows a consistency between the     
cluster masses obtained by traditional dynamical methods       
and gravitational lensing, while the latter is independent of
dynamical state and matter content. Therefore, 
such a cosmological baryon fraction can be used for the determination
of the mean mass density of the universe. In
combination with the BBN prediction an open universe 
of $\Omega_m<0.3$ is thus preferred. It turns out that the solution to
the so-called baryon crisis is either to modify the standard BBN
model or inflationary model, or to introduce a nonzero
cosmological constant into physics/astrophysics.

The increasing tendency of the observed cluster                    
baryon fraction  to $\sim30\%$ with scales (Figure~\ref{fig-1})               
has brought about not only                                         
the baryon crisis, if we live in a flat universe and
if we do not intend to accept a nonzero cosmological constant,
but also a mystery that a greater fraction of dark matter exists               at small scales than at  large scales.  Future cosmological
study of structure formation should take this fact into account.
Meanwhile, we should be aware that the fraction of baryonic
matter in the universe is not small at all: $\sim30\%$ of the
matter of the universe is actually visible !

\acknowledgments
I thank my collaborators Prof. Li-Zhi Fang, Tzihong Chiueh and
Yan-Jie Xue for many simulating discussions. 
This work was supported by the National Science Council of Taiwan, 
under Grant No. NSC87-2816-M008-010L,      
and the National Science Foundation of China, under Grant No. 19725311.

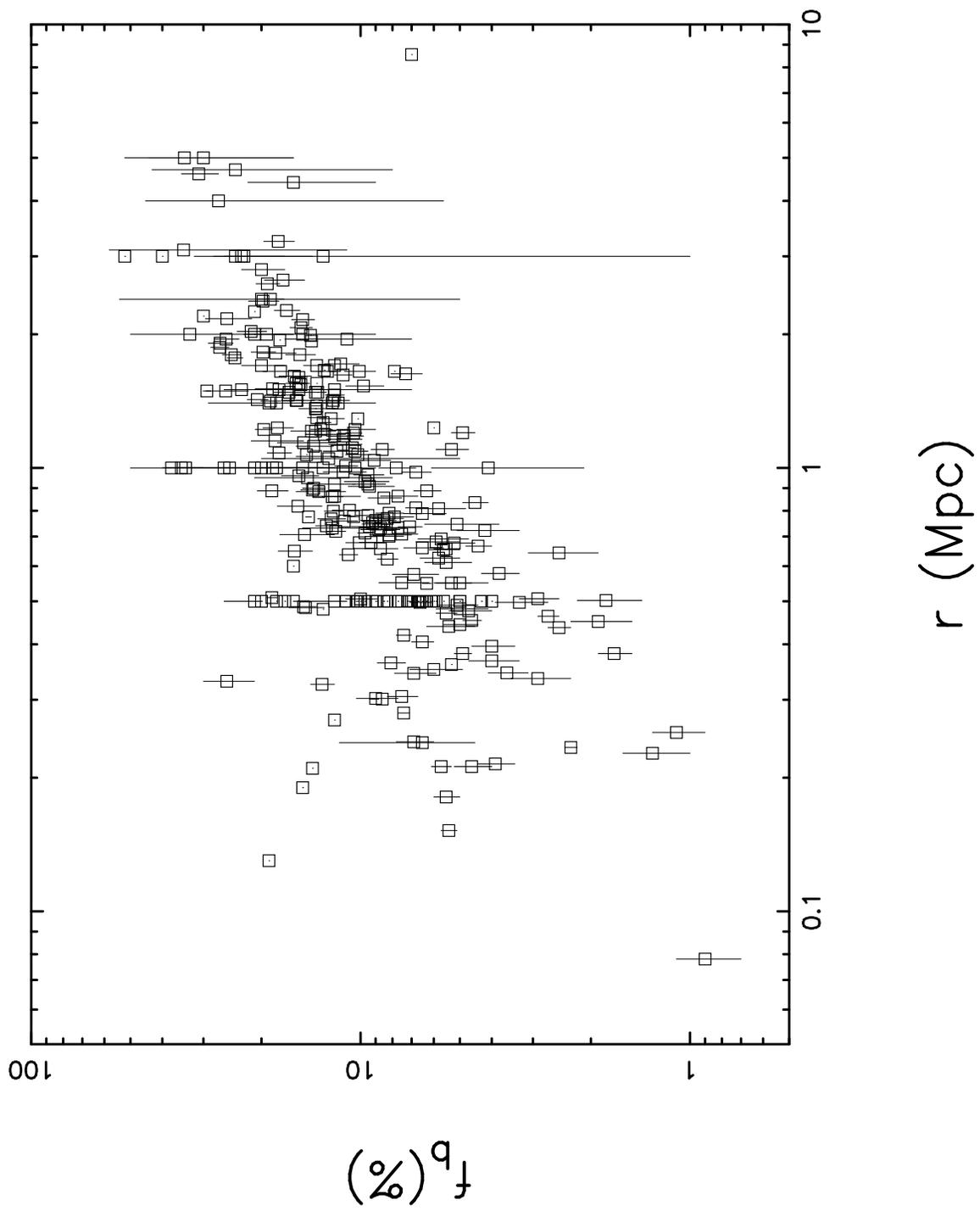
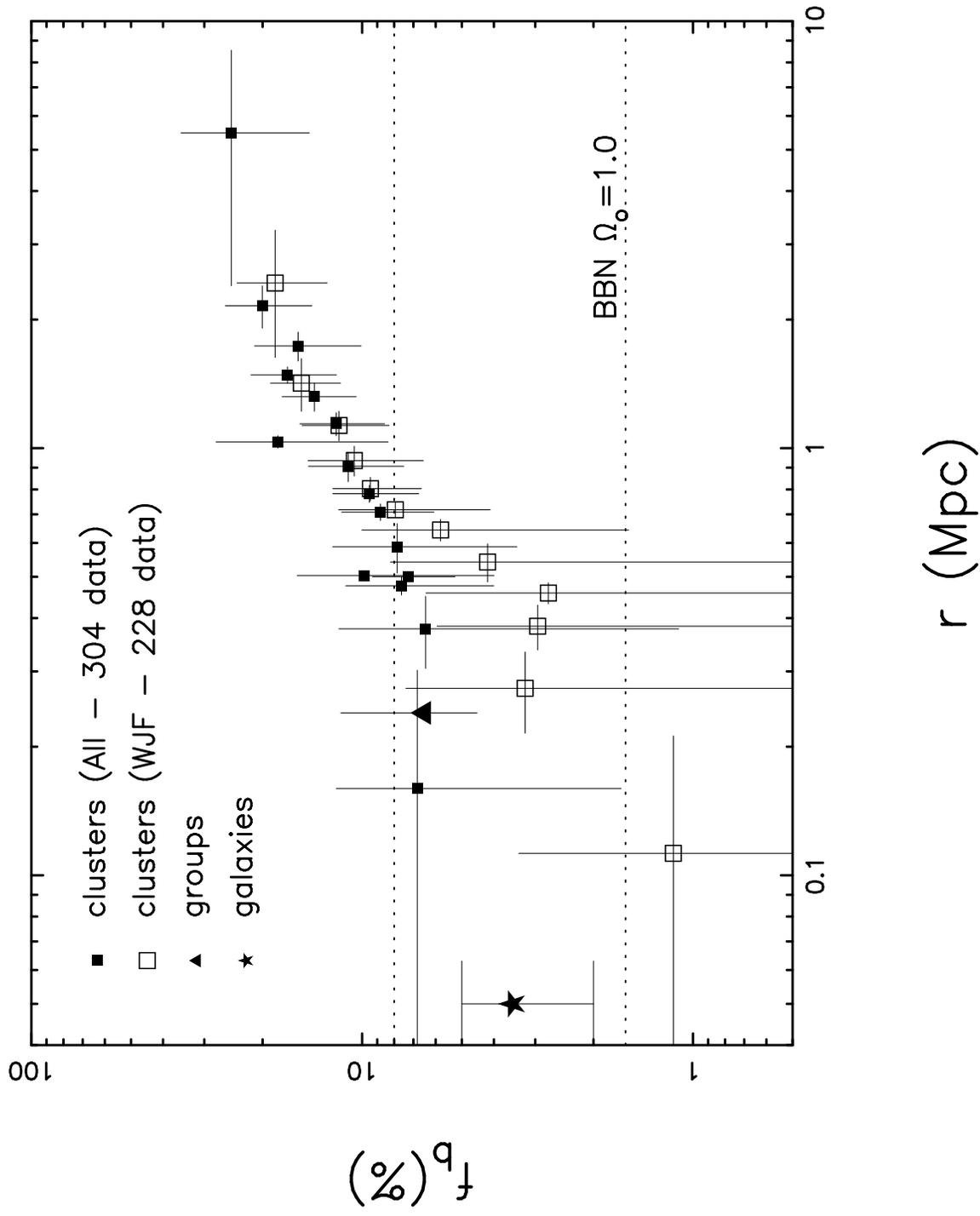
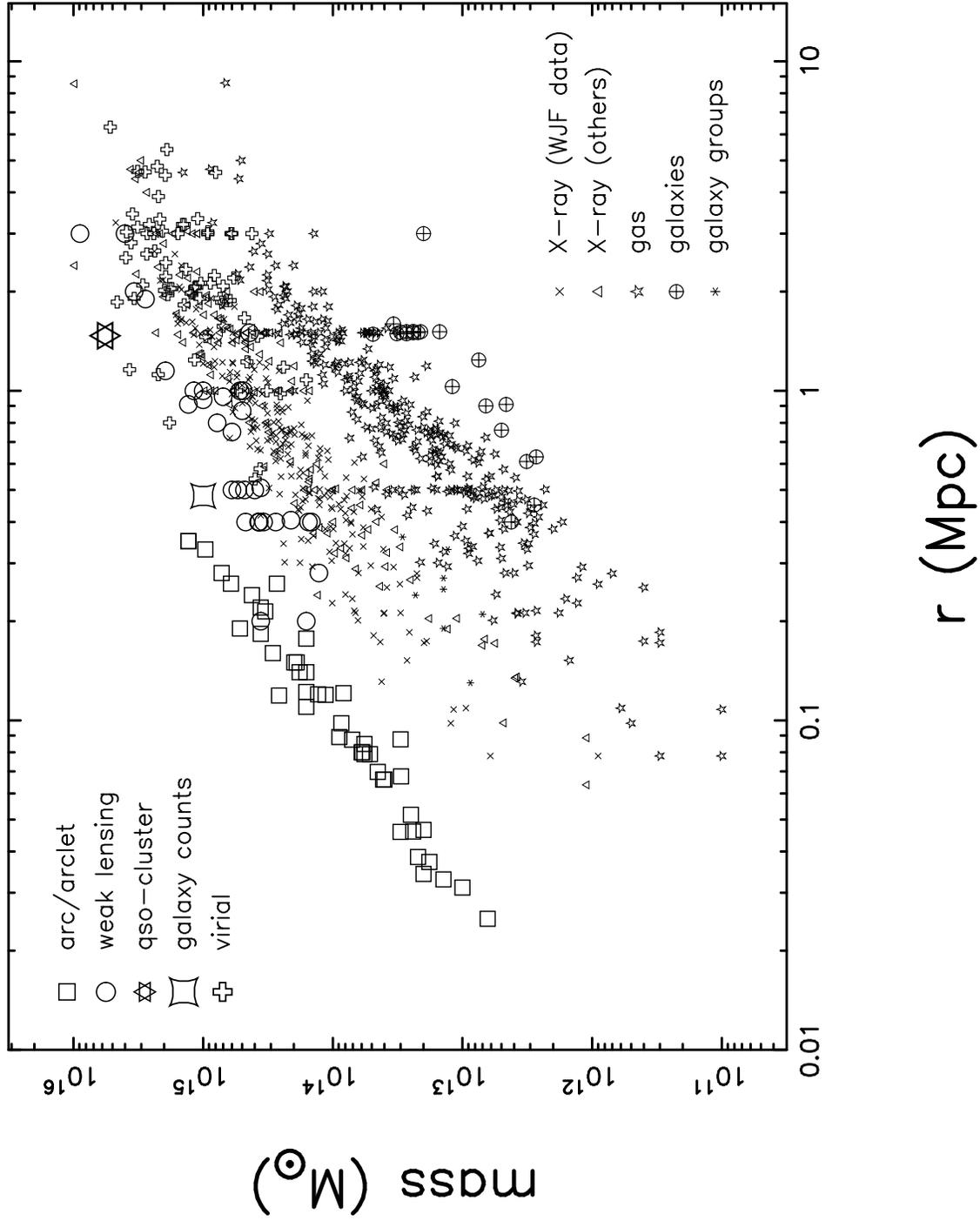
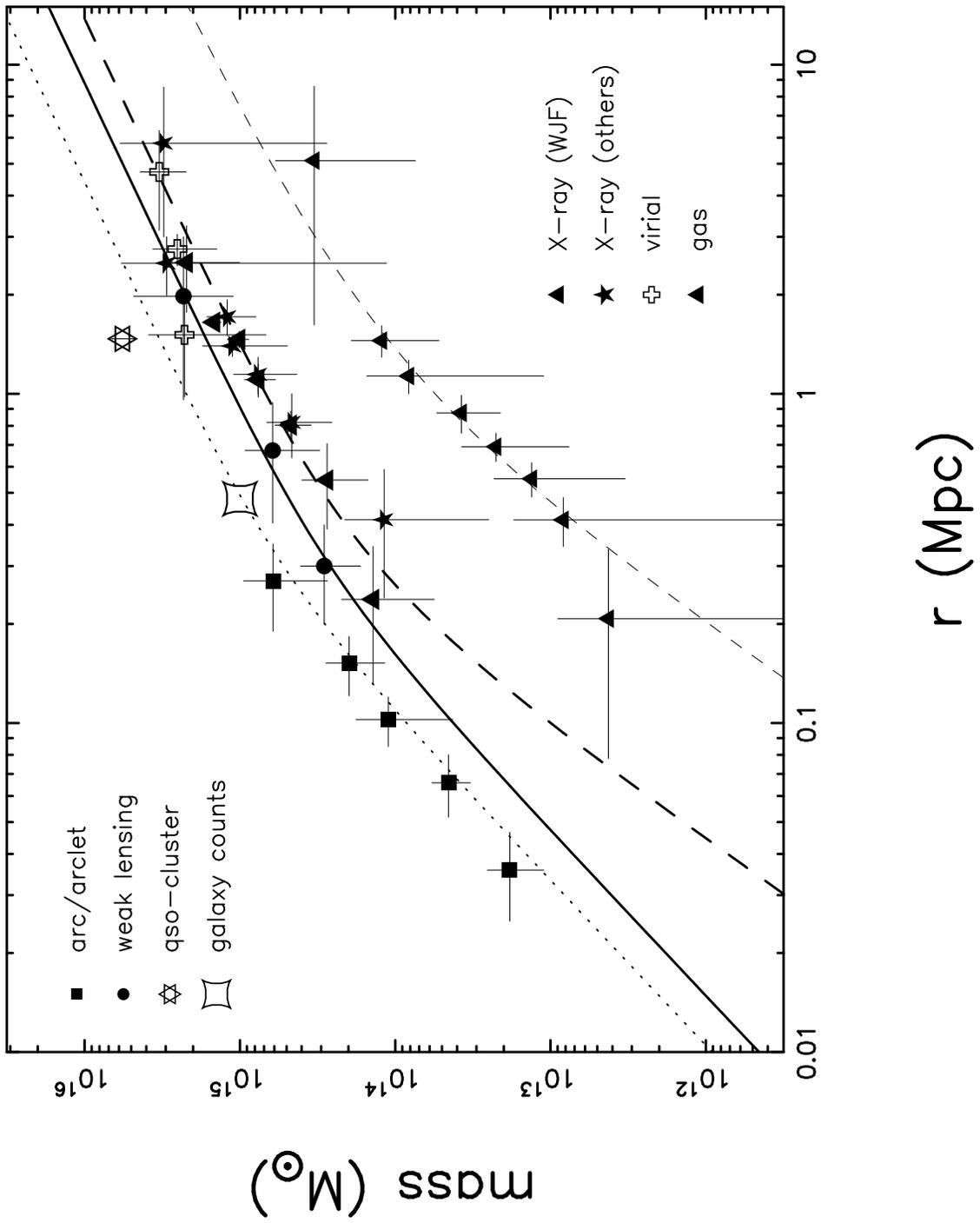
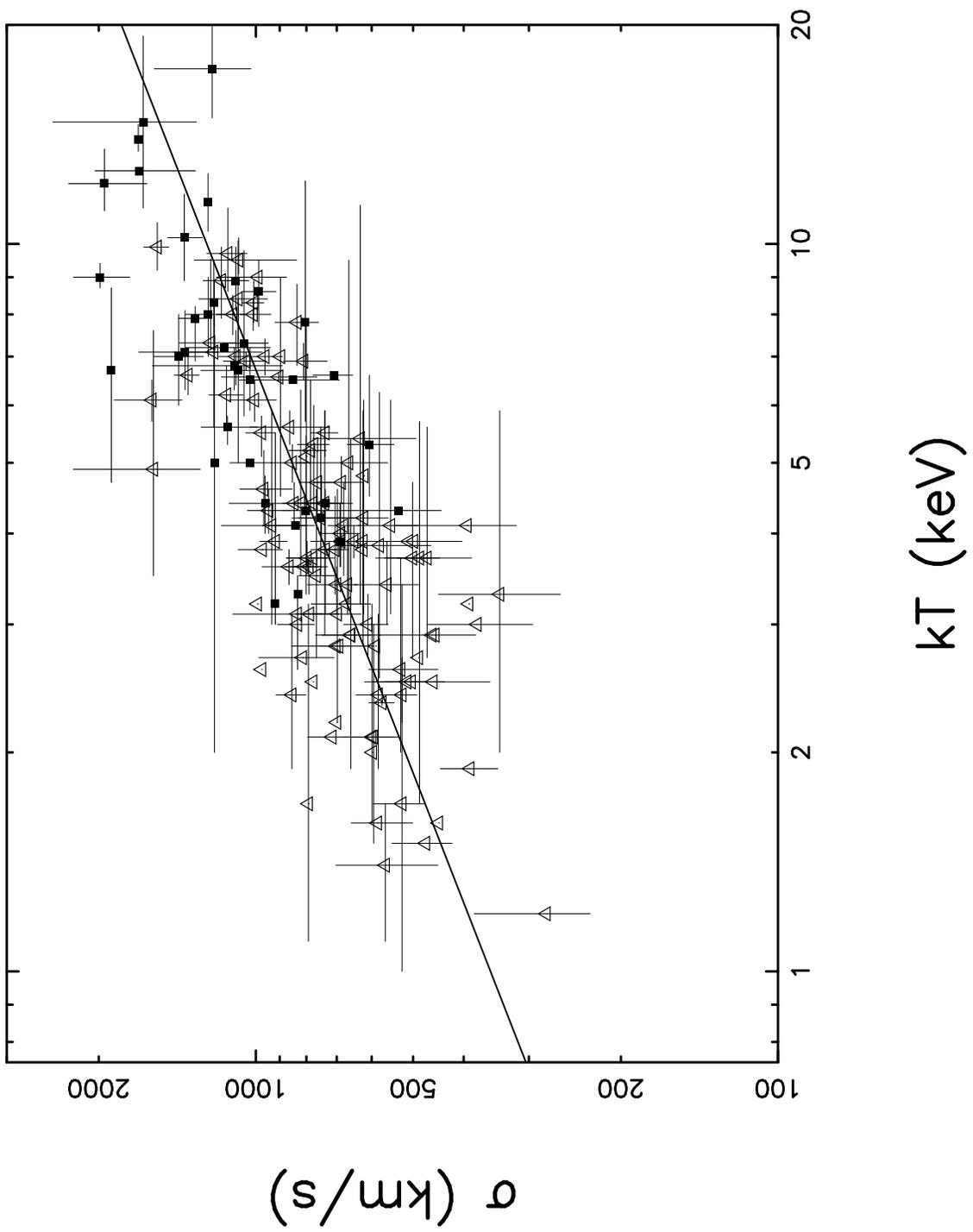
\end{document}